# Wide field-of-view and large depth-of-field metalenses

Louis Martin-Monier[1], Zhaoyi Li[1*], Fan Yang[1,2], Mikhail Shalaginov[1], Luigi Ranno[1], Jia Xu Brian Sia[1], Tian Gu[1,2*], and Juejun Hu[1,2]

[1]Department of Materials Science & Engineering, Massachusetts Institute of Technology, Cambridge, Massachusetts, USA
[2]2Pi Inc., Cambridge, Massachusetts, USA

*zhaoyili@mit.edu, gutian@mit.edu

**Abstract**

The ability to visualize both macroscopic and microscopic features over an extended field of view is essential for endoscopic imaging and other applications ranging from machine vision to microscopy. However, miniaturizing endoscopes introduces inherent trade-offs between size and optical performance, including field-of-view (FOV), depth-of-field (DOF), and resolution. These constraints limit the use of microendoscopes in clinical settings such as early cancer detection within narrow, hard-to-access anatomical regions, including the lung, ovaries, and pancreas. State-of-the-art microendoscopes typically rely on microlens assemblies that increase both cost and size. Their large f-numbers also hinder the collection of high-resolution information from live tissue. In this work, we present two compact metalens designs that provide wide FOV, extended DOF, and high resolution, enabled by custom-tailored point spread functions (PSFs). The devices achieve a full 172° FOV, an extended DOF from 0.4 mm to beyond 300 mm, and a resolution of 30 line pairs per millimeter, all within a 1 mm × 1 mm × 0.2 mm footprint. A key advantage of our approach is the ability to transition seamlessly between low and high magnification without mechanical refocusing. Final images are reconstructed through backend deconvolution, highlighting the potential of hybrid imaging systems that integrate computational techniques with flat-optics components.

**Introduction**

Microendoscopes play a critical role in modern clinical diagnostics by enabling in vivo imaging within narrow and otherwise inaccessible anatomical regions. Their miniaturized form factor makes them particularly well-suited for early detection and monitoring of diseases in organs such as the lung, ovaries, and pancreas, where traditional endoscopic tools are either too large or lack sufficient resolution. These applications are of particular importance in the context of early-stage cancer diagnostics, where timely identification of pathological changes can significantly improve patient outcomes. In addition to oncological use, microendoscopes are increasingly employed in minimally invasive procedures across gastroenterology, gynecology, and pulmonology, where real-time visualization is essential for biopsy guidance, tissue characterization, and surgical navigation.

A number of miniaturized endoscopes have been developed to improve access to hard-to-reach anatomical regions while minimizing patient trauma. Techniques such as fiber-bundle imaging[1–3], spectrally encoded endoscopy[4,5], scanning fiber endoscopy[6], and chip-on-tip devices[7,8] have enabled progress toward this goal. The primary optical performance metrics for microendoscopes include FOV, DOF, and resolution. Achieving a wide FOV typically involves the use of cascaded lens systems, as in traditional fisheye optics. However, this approach increases the axial footprint of the optics, and any additional axial length imposes a lower bound on the minimum bending radius of the endoscope, thereby limiting maneuverability in surgical environments. An alternative strategy is to employ two independent optical assemblies, each covering a portion of the FOV, but this too comes at the cost of increased device size.

In addition, wide FOVs are typically realized at high f-numbers, which reduce numerical aperture and restrict imaging at high magnification. The ability to observe tissue at increased magnification is a critical requirement in endoscopy, particularly for the early detection of minute lesions. For instance, magnification endoscopy is valuable in screening patients with Barrett's esophagus or colorectal cancer and can be further enhanced with topical stains (magnification chromoendoscopy). High-magnification imaging requires optics with small f-numbers, which in turn demand a large DOF to mitigate the effects of *in vivo* tissue motion. While some commercial systems have incorporated switchable-focus objective lenses to address this challenge (e.g., Olympus EVIS X1 and ITRI's EDoF endoscope), these solutions add size and complexity. In principle, a large DOF at low f-number could allow seamless magnification without refocusing, but current systems remain limited in DOF and fail to provide this capability in a compact form.

Optical metasurfaces have emerged as a powerful platform to overcome the longstanding trade-off in miniaturized optical systems. Composed of subwavelength scatterers that impart spatially varying phase delays, metasurfaces enable precise wavefront control in ultra-thin form factors. In the context of microendoscopy, metasurfaces offer significant advantages over traditional refractive optics by drastically reducing optical stack height, correcting aberrations over wide angles, and extending DOF without the need for mechanical focus adjustment[9]. A number of experimental studies have demonstrated metasurface-enabled microendoscopes across various imaging modalities, including optical coherence tomography[10], capsule endoscopy[11], fiber scanning probes[12], and fiber bundle imaging[13–15]. Table 1 summarizes key performance metrics reported in recent metasurface-based endoscopic imaging studies.

Here, we demonstrate a doublet metasurface architecture (Fig. 1a) combined with backend image processing to simultaneously achieve imaging with a near-180° FOV and extended DOF reaching submillimeter object distances, while maintaining high resolution. In this architecture, the first metasurface imposes a uniform PSF modulation on incoming light from all directions.

The second metasurface performs the primary focusing operation, forming an image on the sensor plane, a configuration reminiscent of our previously reported fisheye metalens design[16–18]. A key feature of the doublet architecture is that the resulting PSF remains nearly invariant with respect to the angle of incidence. This angular invariance greatly simplifies computational post-processing by enabling angle-independent deconvolution, which is efficient enough to support real-time, video-rate imaging. We implemented two design variants based on this shared framework. In one, the front metasurface encodes a cubic phase profile, known to produce a PSF that is invariant to defocus[19]. In the other, the front metasurface applies a logarithmic-asphere phase profile, which generates a focal length that varies continuously with radial position, ensuring that at least one annular region of the lens is always in focus[20]. Benchmarking the performance of these designs against prior work (Table 1) highlights the potential of such compact and scalable assemblies for next-generation microendoscopic imaging systems.

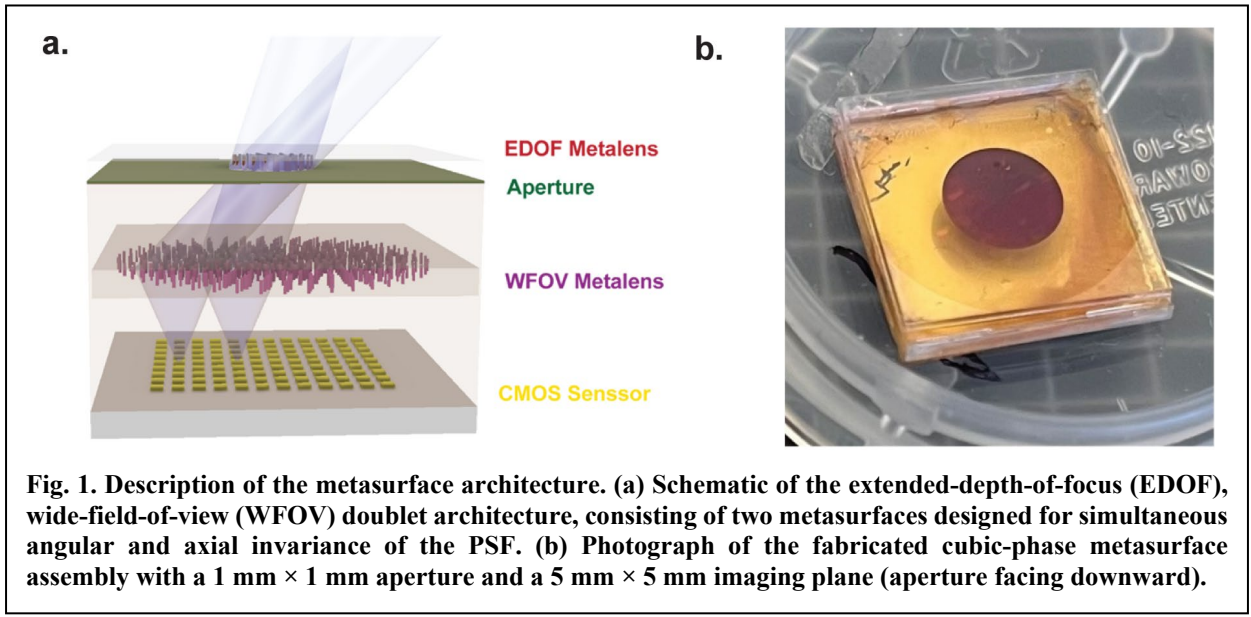

**Fig. 1. Description of the metasurface architecture.** (a) Schematic of the extended-depth-of-focus (EDOF), wide-field-of-view (WFOV) doublet architecture, consisting of two metasurfaces designed for simultaneous angular and axial invariance of the PSF. (b) Photograph of the fabricated cubic-phase metasurface assembly with a 1 mm × 1 mm aperture and a 5 mm × 5 mm imaging plane (aperture facing downward).

**Results**

The phase function of the cubic-phase metasurface is defined as:

$$\phi(r) = \frac{\alpha}{R^3}(x^3 + y^3) \qquad (1)$$

where $R$ is the aperture radius and $\alpha$ is a design constant, chosen here as $1000\pi$. $\alpha$ modulates the strength of the phase gradient and is chosen here to maximize the depth of focus. The phase profile of the second metasurface was initially determined using an analytical solution[17] and subsequently fine-tuned through ray tracing simulations in Zemax OpticStudio®. The diameters of the first and second metasurfaces are 1 mm and 5 mm, respectively, spaced by 2.3 mm. The back focal length of the doublet is 2 mm, corresponding to an f-number of 2. The corresponding PSFs, modelled using Zemax OpticStudio®, are shown in Fig. 2a for various incidence angles and object distances. For comparison, Fig. 2b presents the experimentally measured PSFs from a collimated point source under oblique incidence up to 80° (see Figure SI 1 for performance up to 86°). Both simulation and experimental results confirm that the PSF remains nearly invariant with respect to the direction of incident light, demonstrating angle-independence over a wide FOV.

We next performed imaging tests with the cubic-phase metasurface doublet using a 1951 USAF resolution test chart. To extract high-quality images from the raw data captured by the metalens, the output must be deconvolved with the system's PSF. Owing to the PSF's angular invariance across the FOV, a single PSF and a fixed set of parameters are sufficient to deconvolve all captured images with acceptable accuracy. This feature significantly reduces computational latency and simplifies post-processing.

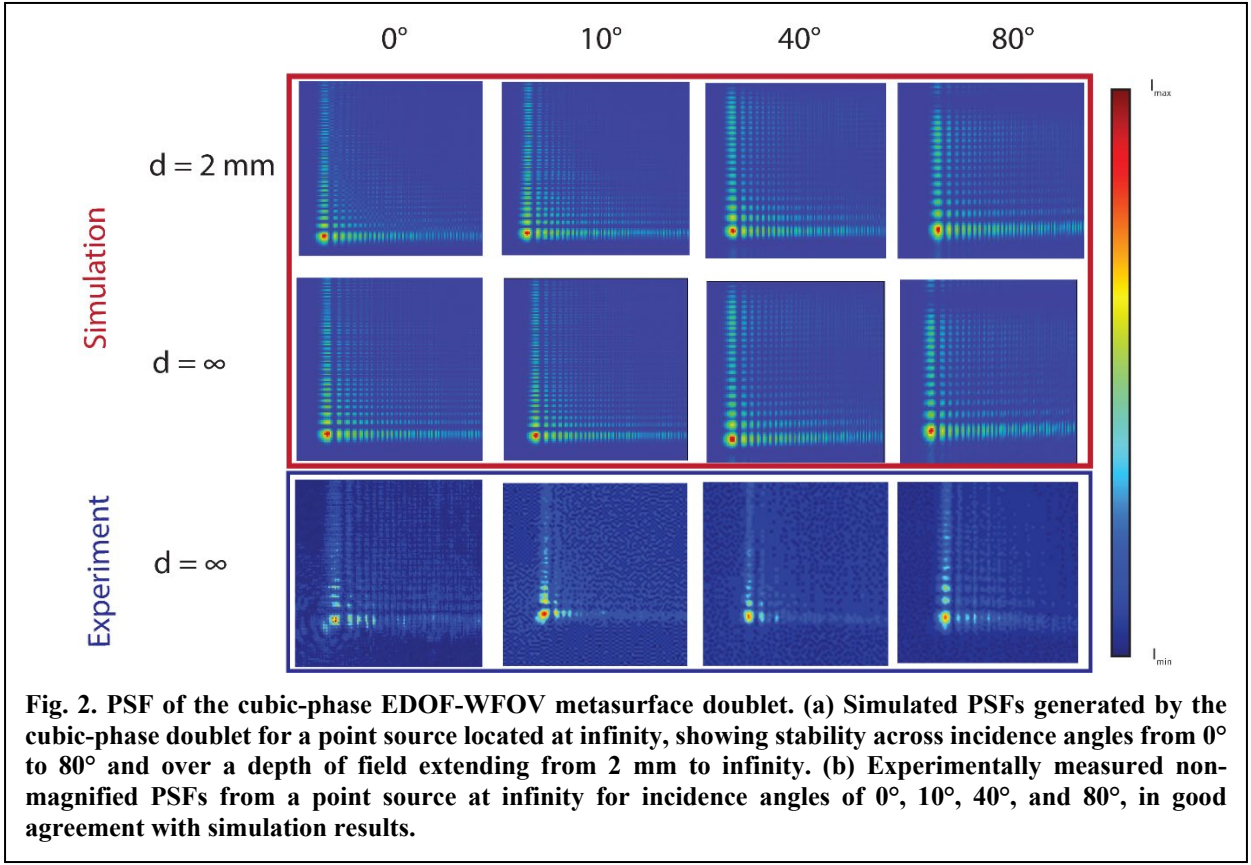

**Fig. 2. PSF of the cubic-phase EDOF-WFOV metasurface doublet. (a) Simulated PSFs generated by the cubic-phase doublet for a point source located at infinity, showing stability across incidence angles from 0° to 80° and over a depth of field extending from 2 mm to infinity. (b) Experimentally measured non-magnified PSFs from a point source at infinity for incidence angles of 0°, 10°, 40°, and 80°, in good agreement with simulation results.**

The deconvolution was performed using the Wiener filter algorithm, expressed as:

$$G(u,v) = \frac{H^*(u,v)}{|H(u,v)|^2 + \frac{P_u(u,v)}{P_*(u,v)}} \quad (2)$$

where $G(u,v)$ is the Wiener filter, $H(u,v)$ is the Fourier transform of the PSF, $P_u(u,v)$ is the power spectrum of the signal, and $P_n(u,v)$ is the power spectrum of the noise. The ratio $P_u/P_s$ is inversely related to the signal-to-noise ratio (SNR). Using the Wiener filter, the reconstructed scene $\hat{S}(u,v)$ is obtained from the measured image $X(u,v)$ via:

$$\hat{S}(u,v) = G(u,v)X(u,v) \quad (3)$$

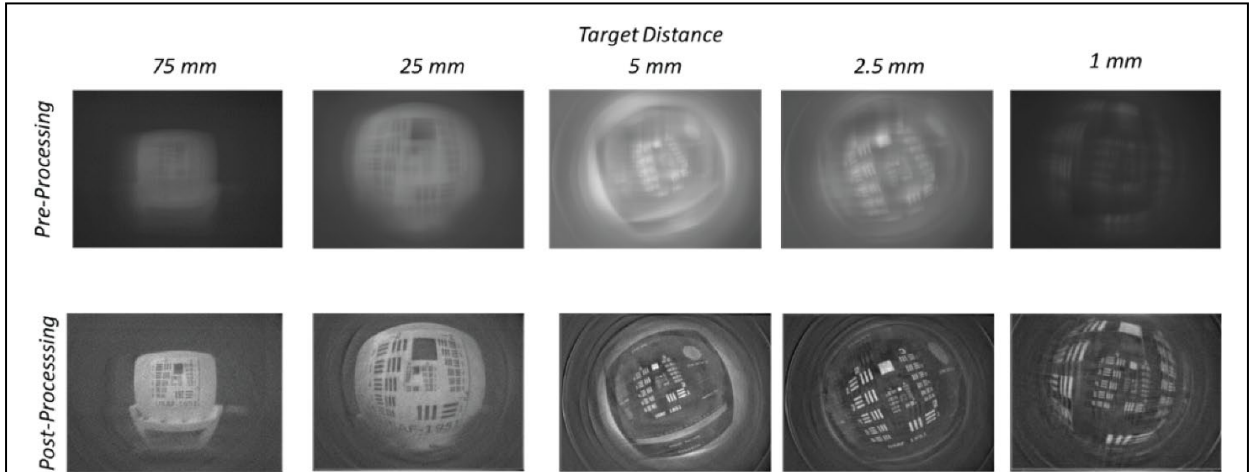

**Fig. 3.** Imaging results of the 1951 USAF resolution target captured using the cubic-phase metasurface assembly. The top row shows raw images acquired before post-processing, while the bottom row presents corresponding images reconstructed through deconvolution. The target distance decreases from left to right, ranging from 75 mm to 1 mm.

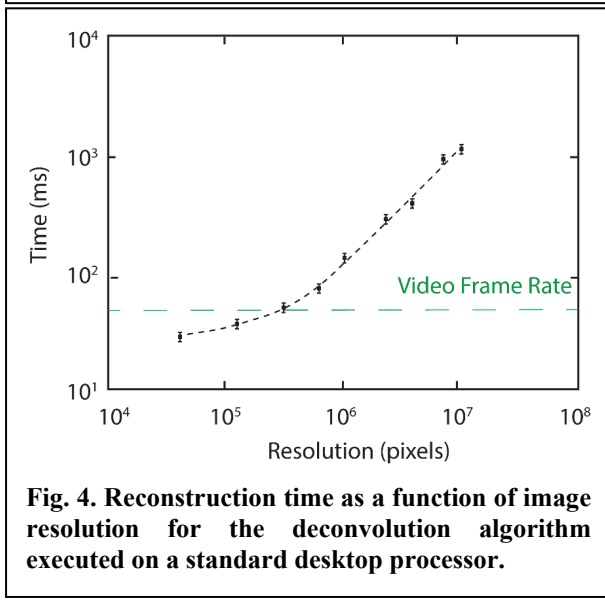

**Fig. 4.** Reconstruction time as a function of image resolution for the deconvolution algorithm executed on a standard desktop processor.

Figure 3 presents a comparison between unprocessed raw images and images deconvolved using a constant single PSF, taken as the simulated PSF at normal incidence. Following deconvolution, the image contrasts were normalized relative to their respective maxima to ensure visual continuity suitable for video display. The processed images are able to resolve a wide range of USAF 1951 groups over a broad set of distances, ranging from Group -2 to Group 4 element 6 (e.g., 30 lp/mm resolution) over a depth range spanning from 75 mm to 1 mm.

Using a single point spread function (PSF) to deconvolve an image assumes shift-invariance of the optical system and offers a major computational advantage over space-variant or multi-PSF deconvolution. In the single-PSF case, the operation reduces to two fast Fourier transforms (FFTs) and one element-wise division, with complexity scaling as $O(N^2 \log N)$ for an image of N×N pixels and memory requirements that scale with image size, e.g. $O(N^2)$. In contrast, when the PSF varies across the field and the image is divided into tiles, each subregion must be independently padded, transformed, and blended using overlap-add or overlap-save schemes. Considering a basic 3 × 3 image tiling, this leads to roughly 5–10 × higher processing time and 2–3 × larger memory footprint due to repeated computations on overlapping margins.

Figure 4 summarizes the post-processing latency of the current algorithm as a function of image resolution, measured on standard desktop-class CPU hardware (see Methods). It requires approximately 1.1 seconds to process a full-resolution image of 3856 × 2764 pixels (about $10^7$ pixels), and about 44 milliseconds for a 500 × 500 pixel image ($2.5 \times 10^5$ pixels). Notably, this resolution aligns with the pixel count of state-of-the-art chip-on-tip cameras, which typically integrate up to several $10^5$ pixels. The presented post-processing pipeline is therefore compatible with video-rate imaging, enabling real-time visualization for operating surgeons. We further note that several approaches exist to accelerate or simplify the post-processing pipeline, including deep learning–based deconvolution methods[21], blind deconvolution algorithms (that do not require prior knowledge of the PSF)[22], and advanced denoising techniques[23]. While the implementation of such methods is beyond the scope of this work, they underscore the significant potential for enhancing our hybrid imaging framework, which combines flat optical elements with computational image reconstruction.

Our second design employs a logarithmic-asphere phase profile that produces a uniform intensity distribution along a continuous line of foci under a collimated incident beam. The phase function is given by:

$$\phi(r) = \frac{2\pi}{\lambda} \int_0^r \frac{r' \mathrm{d}r'}{\left\{r'^2 + \left[s_1 + (s_2 - s_1)\left(\frac{r'}{R}\right)^2\right]^2\right\}^{1/2}} \quad (4)$$

where $r' = \sqrt{x^2 + y^2}$. For this design, we set $s_1$ = 0.3 mm to $s_2$ = 1 mm, achieving an extended depth of focus at a high numerical aperture corresponding to an f-number of 1.5. Figure 5 presents a comparison between the simulated and experimentally measured PSFs of the logarithmic-asphere design, showing excellent agreement across the field.

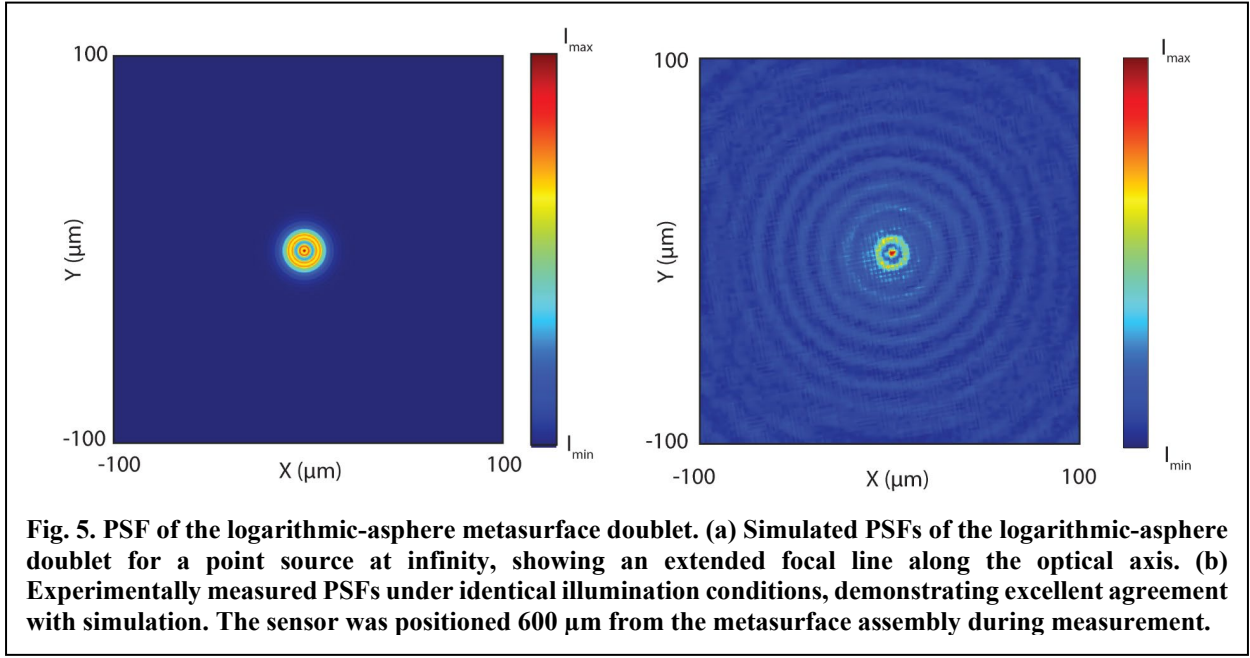

**Fig. 5. PSF of the logarithmic-asphere metasurface doublet. (a) Simulated PSFs of the logarithmic-asphere doublet for a point source at infinity, showing an extended focal line along the optical axis. (b) Experimentally measured PSFs under identical illumination conditions, demonstrating excellent agreement with simulation. The sensor was positioned 600 μm from the metasurface assembly during measurement.**

Compared to the cubic-phase PSF, the logarithmic-asphere PSF offers two key advantages: (i) it is rotationally symmetric, which reduces spatially varying noise artifacts in the reconstructed image, and (ii) it produces a final image that is directly interpretable, often eliminating the need for explicit deconvolution. As shown in Fig. 6, the metasurface doublet with a 200 μm aperture is capable of resolving line features separated by approximately 4 μm (USAF Group 6, Element 6), corresponding to an estimated resolution limit of 115 lp/mm. The focusing performance remains stable over an axial range extending from 400 μm to 1.5 mm. It is worth noting that these images were captured using a CMOS sensor with a pixel size of 1.67 μm × 1.67 μm, and that the illuminated region shown in Fig. 6 spans approximately 120 × 120 pixels. Because illuminating a nearby target was impractical, a collimated beam and a USAF 1951 test chart were used, limiting the field of view due to the setup rather than the lens itself. The system operates through angular magnification, which increases as the object moves closer, enabling finer spatial detail until the detector's Nyquist limit is reached. Beyond this point, pixel sampling, rather than optical performance, becomes the dominant constraint, as further magnification only enlarges the image without adding resolvable information.

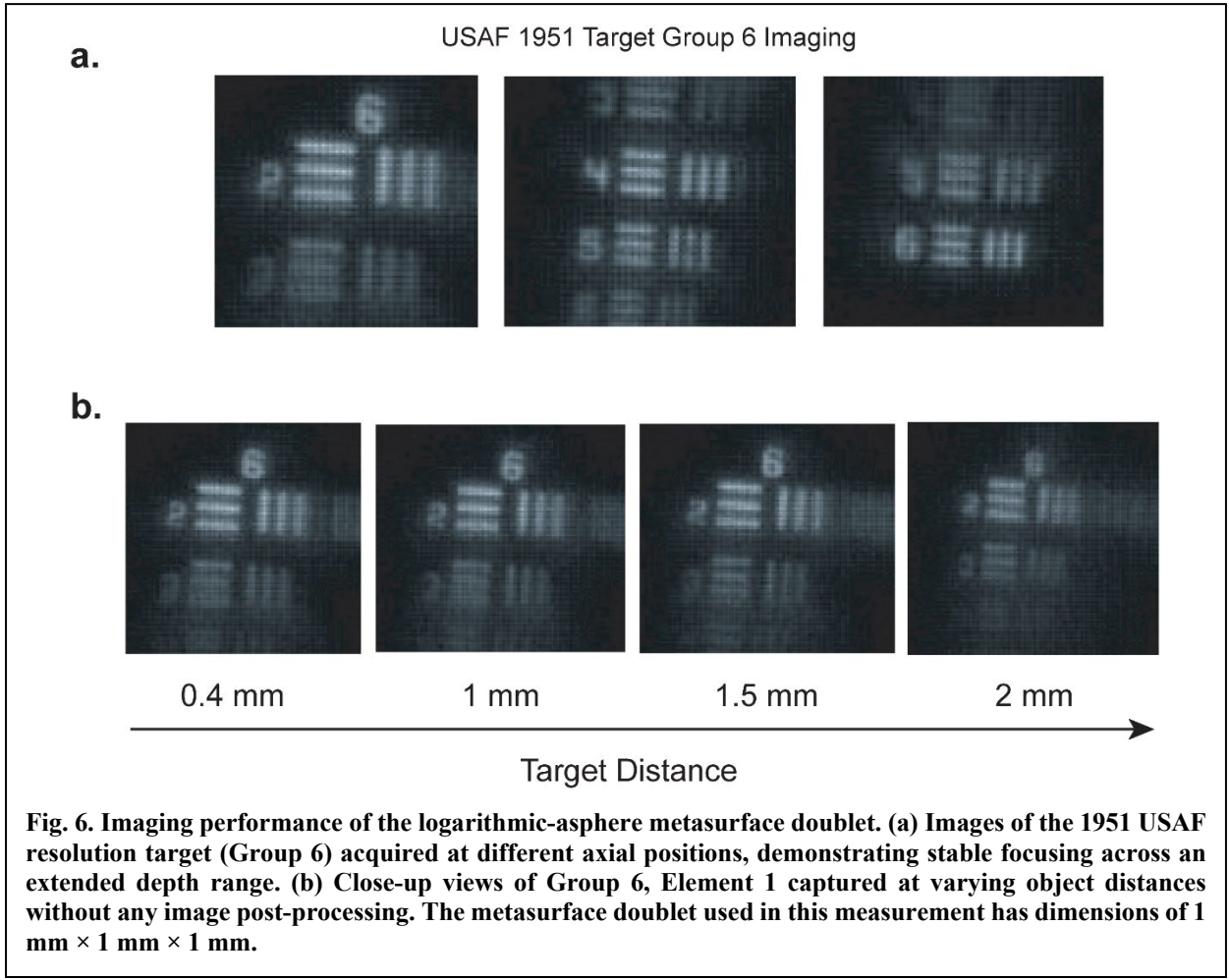

Fig. 6. Imaging performance of the logarithmic-asphere metasurface doublet. (a) Images of the 1951 USAF resolution target (Group 6) acquired at different axial positions, demonstrating stable focusing across an extended depth range. (b) Close-up views of Group 6, Element 1 captured at varying object distances without any image post-processing. The metasurface doublet used in this measurement has dimensions of 1 mm × 1 mm × 1 mm.

**Conclusions**

In this work, we presented two compact metasurface doublet designs that simultaneously achieve wide FOV, extended DOF, and high spatial resolution, addressing longstanding trade-offs in microendoscopic optics. Leveraging custom-tailored PSFs encoded into meta-optical elements, our designs demonstrate near-180° FOV, DOF extending from submillimeter to far-field infinity, and resolution approaching 115 lp/mm within a sub-millimeter-scale form factor.

We explored two PSF modulation strategies: a cubic-phase profile enabling robust post-processed image recovery via angle-invariant deconvolution, and a logarithmic-asphere profile yielding rotationally symmetric PSFs and interpretable images without computational restoration. Both approaches were validated through simulations and experimental measurements, showing excellent agreement in PSF and imaging performance. Notably, the angular invariance of the PSFs enables efficient deconvolution with a single kernel, supporting video-rate operation compatible with current chip-on-tip sensors.

This study demonstrates the promise of combining metasurface optics with computational imaging pipelines to realize high-performance, miniaturized imaging systems. Our results open pathways for ultra-compact, real-time endoscopic systems suitable for clinical applications, including early cancer diagnostics and minimally invasive procedures, where imaging flexibility, resolution, and size are critical.

**Table 1. Representative experimental demonstrations of metasurface-based endoscopic imaging. N/A indicates that the corresponding information was not provided in the referenced publication.**

| | Imaging modality | Wavelength (nm) | Aperture size (mm) | FOV (degree) | DOF (mm) | f-number | Resolution |
|---|---|---|---|---|---|---|---|
| *Nanophotonics* **13**, 4417 (2024)[11] | Direct imaging | 940 | 0.26 | 165 | N/A | 2.34 | 144 lp/mm |
| *Comms. Eng.* **4**, 53 (2025)[12] | Fiber scanning | 643, 532, 444 | 0.68 | 70 | N/A | 0.59 | 0.5° |
| *eLight* **3**, 13 (2023)[13] | Direct imaging | Broadband visible | 1 | 22.5 | 7 – 40 | 2 | 11 lp/mm |
| *Photonics. Res.* **9**, 106 (2021)[14] | Direct imaging | 525 | 0.4 | 35 | N/A | 1.1 | 228 lp/mm |
| *Light Sci. Appl.* **13**, 305 (2024)[15] | Phase imaging | Broadband visible | 0.5 | 28 | N/A | ~ 2 | $0.2\pi$ |
| *Nat. Commun.* **13**, 4183 (2022)[24] | Fiber scanning | 1250 – 1650 | 0.1 | 15 | N/A | 2.4 | 102 lp/mm |
| This work: cubic phase metalens | Computational imaging | 940 | 1 | 172 | 1 – 300 | 2 | 30 lp/mm |
| This work: logarithmic-asphere metalens | Direct imaging | 940 | 0.2 | 172 | 0.4 – 2 | 2 | 115 lp/mm |

## Methods

<u>Metasurface Doublet Fabrication:</u> Fabrication began with a fused silica substrate (University Wafer, Inc.). A 975-nm-thick amorphous silicon (a-Si) film was deposited by plasma-enhanced chemical vapor deposition (SAMCO, Inc.) at 270 °C. The metasurface pattern was defined using electron beam lithography (HS-50, STS-Elionix) operated at 50 kV and 10 nA, followed by reactive ion etching (Pegasus, SPTS Technologies). A 320-nm-thick ma-N 2403 resist was spin-coated at 3000 rpm for 1 min and soft-baked at 90 °C for 2 min. Prior to coating, the sample was treated with HMDS vapor priming to improve resist adhesion. To mitigate charging effects during exposure, a water-soluble conductive polymer (ESpacer 300Z, Showa Denko America, Inc.) was applied atop the resist. After exposure, the resist was developed in a TMAH-based developer (AZ 726 MIF, Micromaterials). Reactive ion etching was carried out using alternating $SF_6$ and $C_4F_8$ gas cycles. The remaining resist was removed via oxygen plasma. A thin layer of SU-8 (Kayaku Advanced Materials Inc., Japan) was then spin-coated and baked to encapsulate and protect the meta-atom structures. The EDOF and WFOV metasurfaces were fabricated on separate substrates. Passive alignment and bonding of the two layers were performed using reference marks and an MRSI-A-L Die Bonder, ensuring accurate registration of the metasurface doublet.

<u>Optical Simulation</u>: The optical performance of the metasurface assembly was evaluated using a custom simulation framework based on the Kirchhoff diffraction integral. The model first optimized the telecentricity of the WFOV phase mask using a single-aperture, single-layer configuration. Once the telecentric condition was established, the complete doublet assembly was simulated by positioning the EDOF phase profile at the first layer's aperture to ensure PSF stability across the field angles. The simulations were cross-validated in sequential mode using Zemax OpticStudio® (Ansys Inc.). Both metalenses were modeled as phase masks: the WFOV element was represented by a 24th-order symmetric polynomial, and the EDOF element by a cubic polynomial in *x* and *y*. Each mask was implemented on a silica substrate, bonded with an index-matching adhesive. The polynomial coefficients were optimized to maximize the Strehl ratio across incidence angles from 0° to 85°. The PSFs and modulation transfer functions (MTFs) were computed using the Fast Fourier Transform module in Zemax OpticStudio®. The meta-atom phase responses were extracted via rigorous coupled-wave analysis (RCWA, RETICOLO[25]) at the design wavelength of 940 nm under normal incidence and periodic boundary conditions.

<u>Optical Characterization</u>: Optical measurements were conducted using a 940 nm continuous-wave laser diode module (LQC940-90E, Newport Inc.) as the illumination source. The laser beam was collimated through a beam expander (ZBE22, Thorlabs Inc.) and directed onto the metasurface doublet mounted on a three-axis translation stage with an integrated two-axis tilt adjustment for fine angular alignment. The focused light was collected using a 50× microscope objective and projected onto a 10 MP monochrome CMOS camera module (MT9J001, Arducam) with a 1/2.3" Aptina sensor (3856 × 2764 pixels 7.5 fps, 1.67 μm pixels). The camera and objective were mounted on an independent motorized stage, allowing precise axial scans to locate the focal plane. The intensity distribution at each plane was recorded without filtering to capture the full point spread function (PSF) response. For imaging experiments, a 1951 USAF resolution target was positioned at various object distances to evaluate depth-of-focus (DOF) and resolution. Deconvolution was performed using a Wiener filter algorithm implemented on an Intel® Core™ i7-7700 CPU (3.60 GHz, 4 cores, 8 threads).

**Acknowledgments**

This work was sponsored by the DARPA ENVision program. L.M. acknowledges funding support provided by the Swiss National Science Foundation (P500PT_203222). The views, opinions and/or findings expressed are those of the authors and should not be interpreted as representing the official views or policies of the Department of Defense or the U.S. Government. Z.L. and J.H. acknowledges funding support provided by the Ministry of Trade, Industry, and Energy (MOTIE), Korea, under the Global Industrial Technology Cooperation Center Program (reference number P0028512) supervised by the Korea Institute for Advancement of Technology (KIAT).

**Author contributions**

L.M. performed finite element modelling, developed fabrication protocols, and fabricated and characterized the metasurface optics. S.C. designed and characterized the compound eye micro-metalens array. J.F., L.R., K.P.D., A.U., and J.X.B.S. contributed to device fabrication. Z.L. and H.Z. assisted with metasurface optics testing. J.H. conceived the study. T.G., Y.M.S., A.M., and J.H. supervised and coordinated the research. L.M., S.C., Y.M.S., and J.H. drafted the manuscript. All authors contributed to technical discussions and writing the paper.

**Materials & Correspondence**

Correspondence and requests for materials should be addressed to Zhaoyi Li and Tian Gu.

**Competing financial interests**

A patent based on the technology described herein has been filed and licensed to 2Pi Inc.